\documentclass{llncs}
\usepackage{bm}
\usepackage{graphicx}
\usepackage{algorithm}
\usepackage{algorithmic}
\usepackage{esvect}
\usepackage{amsmath}
\usepackage{placeins}
\title{A Hybrid Recommendation Framework for Enhancing User Engagement in Local News}
\author{Payam Pourashraf\inst{1} \and Bamshad Mobasher\inst{1}}
\institute{
School of Computing, DePaul University, Chicago, IL, USA \\
\email{ppourash@depaul.edu}, \email{mobasher@cs.depaul.edu}
}

\begin{document}
\maketitle

\begin{abstract}
Local news organizations face a pressing need to increase reader engagement amid declining circulation and competition from global media~\cite{abernathy2022}. Personalized news recommender systems offer a promising solution by tailoring content to user interests. However, conventional approaches often focus on user general (global) preferences and may neglect nuanced or eclectic user preferences in the local news context~\cite{abernathy2022}. In this work, we propose a novel hybrid news recommender that integrates local and global preference models to enhance user engagement in local news. Building on previous research that identified the value of localized recommendation models for certain news categories~\cite{pourashraf2022}, our approach combines the strengths of both local and non-local preference predictors within a unified framework. The proposed system adaptively combines recommendations from a \textbf{local model} (specialized for region-specific content) and a \textbf{global model} (capturing general news preferences), using ensemble strategies and multiphase training to balance the two.  We evaluated the hybrid model on two datasets: (1) a large-scale synthetic news dataset based on the Syracuse local newspaper category and locality distributions ~\cite{pourashraf2022}, and (2) a Danish news dataset (EB-NeRD) labeled for local/non-local content using an LLM ~\cite{kruse2024ebnerd}. The results of offline experiments demonstrate that our integrated approach outperforms single-model baselines in prediction accuracy and coverage, suggesting improved personalization that can translate to higher user engagement. The findings have practical implications for news publishers, especially local outlets. Using both community-specific and general user interests, the hybrid recommender can deliver more relevant content to readers, potentially increasing retention and subscription rates. In sum, this work introduces a new direction for news recommender systems that bridges local and global models, offering a scalable solution to revitalize local news consumption through personalized user experiences.

\end{abstract}

\section{Introduction}

Personalized recommendation systems have become integral to digital news platforms, helping users discover relevant content and fostering sustained engagement~\cite{meng2023}. Major media organizations such as \emph{The New York Times} and \emph{BBC} have long used personalized news feeds using techniques ranging from content-based filtering to deep learning~\cite{wu2019,meng2023}. These systems succeed in part by filtering a large collection of articles to match user interests, which has been shown to increase user retention and satisfaction~\cite{meng2023}. However, in the domain of local news, personalization presents unique challenges and opportunities. Local newspapers in the United States have experienced dramatic declines in readership in the past decade, leaving many communities without reliable sources of local journalism~\cite{abernathy2022}. To survive the shift from advertising to subscription-based revenue models, local outlets must provide added value through engagement-rich features such as personalized content recommendations~\cite{abernathy2022}. Recommender system technology is a primary mechanism to deliver this personalization, with the aim of connecting readers with local stories and broader news in line with their interests. The importance of this endeavor is not only commercial but also civic, as effective recommendation can help restore the reader's connection to local news, thus strengthening community information ecosystems.

A key challenge in news recommendation is that user preferences are \textbf{multi-faceted}. Readers of local news sites show interest in community-specific content (e.g., city council updates, local sports) \emph{alongside} and interest in nationally or globally relevant news~\cite{napoli2016}. Traditional recommendation algorithms that model user preferences \emph{global} (aggregating all past behavior) may not capture the situational importance of local content~\cite{abernathy2022}. In contrast, a naive focus on local content \emph{only} could miss articles of broader appeal that also engage the user~\cite{pourashraf2022}. Recent research in recommender systems highlights the benefit of modeling different preference scopes in parallel. For example, some session-based recommenders use neural attention mechanisms to combine short-term session interests with the long-term profile of a user~\cite{wu2019,meng2023}. In the news domain, approaches that incorporate both short-term and long-term preferences have achieved improved accuracy~\cite{meng2023}, and context-aware models that consider factors like time or location (e.g., CROWN~\cite{alabduljabbar2023}) have shown enhanced user engagement. These insights suggest that a \textbf{hybrid approach}—one that can mediate between immediate local interests and overarching global preferences, may produce the most effective recommendations for local news consumers.

In our previous work~\cite{pourashraf2022,pourashraf2023}, we investigated session-based recommendation for local news and introduced a data set that includes user interactions with the website of a local newspaper. That study demonstrated that localized models (trained only on local-content interactions) outperformed global models for certain categories of news. In particular, recommendations within the category \emph{Life \& Culture} - stories with a strong local flavor - were significantly more accurate when using a local-specific model than when using a general news model that did not differentiate content by locality~\cite{pourashraf2022}. This evidence supports the intuition that users' local versus global news preferences can diverge and that capturing the local context can improve the relevance of the recommendation. At the same time, the study found that users do not exclusively consume local content; Integrating global preferences is also necessary for a well-rounded diet of news~\cite{pourashraf2022}. The open question, as identified in our previous conclusions, is how to \textbf{fuse local and global models} into a single recommender system to best serve the user.

This paper extends previous research by developing and evaluating hybrid recommendation models that explicitly integrate local and global user preferences. We posit that a tailored combination of local-focus and global-focus submodels can deliver more personalized and engaging news recommendations than either approach in isolation. In practice, we develop a hybrid recommender architecture that takes advantage of two components: (1) a \emph{Local Model} that learns user interests of a fine-grained nature within local news content, and (2) a \emph{Global Model} that captures the user’s broader interests from all news content. We explore multiple strategies to combine these components, including sequential training (adapting one model based on the other's results) and simultaneous training (jointly optimizing both components), as well as ensemble fusion at prediction time. The research is grounded in an HCI motivation to improve user engagement, but our approach is technical, focusing on the algorithmic fusion of preference models rather than on HCI theory. By situating this work as a continuation of our prior study, we also address its identified limitations: we tested the hybrid models on additional datasets beyond Syracuse and with a significantly larger scale of data, assessing generalizability across locales and languages.

In the remainder of this paper, we first review relevant background and related work on personalized news recommendation and local context modeling. We then detail our experimental design, including the hybrid modeling approaches and the three evaluation scenarios. Through these experiments, we demonstrate the potential of integrating local and global preference models to enhance recommendation accuracy and discuss implications for user engagement in news recommender systems.

\section{Background}

\textbf{Personalized News Recommendation:} Recommender systems have been widely applied in online news to tailor content to users~\cite{meng2023,wu2019}. Approaches range from content-based and collaborative filtering to hybrid methods and deep learning models~\cite{wu2019}. The primary goal is to increase reader engagement and retention by presenting articles that match user interests~\cite{meng2023}. A challenge in news recommendation is the dynamic nature of news: Recency and freshness are crucial, making \emph{session-based recommendation} particularly effective~\cite{petrov2023}. Session-based news recommenders, such as various deep learning architectures, focus on short-term user behavior within a session to capture immediate interests~\cite{wu2019,petrov2023}. However, relying solely on short-term signals can overlook a user’s broader preferences. Therefore, recent work has looked at combining \emph{long-term and short-term preferences}, for example, by incorporating a user’s historical profile alongside clicks in the current session~\cite{meng2023}. These efforts mirror a general understanding in recommender systems: merging multiple perspectives on user preference (e.g. long-term vs. recent or context-specific vs. global) can lead to better predictions~\cite{wu2019,alabduljabbar2023}.

\textbf{Locality in Recommender Systems:} Contextual and location-based information has been studied as a way to improve the relevance of recommendations~\cite{alabduljabbar2023}. In music or mobile application scenarios, factors such as time of day or the geographic location of the user can inform better predictions. In the news domain, using location context has yielded benefits: for instance, context-aware systems such as CROWN integrate user location and temporal patterns to boost accuracy~\cite{alabduljabbar2023}. Despite these advances, the application of personalization in the \emph{local news} sector remains underexplored~\cite{abernathy2022,napoli2016}. Local news recommendation introduces a unique duality: readers are interested in content about their locality, but they also value national/international stories. Prior research has observed that simply treating “locality” as a contextual feature does not automatically improve recommendations for local news audiences~\cite{pourashraf2022,pourashraf2023}. This is because not all local news is equally relevant to every reader, and users of local outlets still engage with a mix of local and non-local stories~\cite{pourashraf2022}. Furthermore, local news often spans diverse categories (e.g., politics, sports, community events) with different user behavior patterns. These observations motivate the need for more nuanced models that can \textbf{distinguish between local and global content preferences} and adjust recommendations accordingly.

\textbf{Previous Work on Local News Recommendations:} Our study is based on the findings of Pourashraf and Mobasher~\cite{pourashraf2022,pourashraf2023}, who conducted empirical investigations of personalized local news recommendations using a session-based framework. In \cite{pourashraf2022}, the authors introduced a data set of user interactions (the Syracuse Local News Dataset) from a local newspaper in Syracuse, NY, spanning four months of clicks. Articles were labeled “local” if they pertained to the state of New York or nearby communities, while articles on national or international topics were considered 'non-local'. The authors evaluated the accuracy of the recommendations under different training conditions: global-only, local-only, and hybrid scenarios. Their results indicated that focusing the model on local content led to higher precision for \emph{local} news, particularly within categories such as \emph{Life \& Culture}~\cite{pourashraf2022}. However, a model trained exclusively on local articles missed relevant general news, confirming that neither purely global nor purely local is sufficient; \textbf{a combination of both} is needed~\cite{pourashraf2022,pourashraf2023}. They concluded by calling for the development of hybrid recommenders that explicitly merge local and global preference models.

In prior work, Pourashraf and Mobasher (2022) adopted a session-based k-Nearest-Neighbor approach (SKNN) as the base recommender. SKNN is a memory-based method that finds the most similar past sessions to the current session and recommends items that appeared in those neighbor sessions~\cite{pourashraf2022}. This choice was well-founded, as SKNN generally outperformed other baseline algorithms on accuracy metrics in their news recommendation experiments~\cite{pourashraf2022}. However, recent advances in sequential recommendation motivate the incorporation of \textbf{SASRec} (Self-Attentive Sequential Recommendation) as an improved base model. SASRec leverages a Transformer-based self-attention architecture to model user item sequences, capturing the relative importance of past items when predicting the next. In particular, SASRec has demonstrated strong performance in sequential (and session-based) recommendation tasks~\cite{kang2018self}. Recent studies even show that a properly tuned SASRec can significantly \textbf{outperform} other modern transformer-based models (e.g. BERT4Rec) under comparable conditions~\cite{zhang2021sbert4rec}. By adding SASRec to our framework, we aim to capitalize on its state-of-the-art sequential modeling capabilities and potentially boost recommendation accuracy.
\section{Proposed Fusion Models}

\subsection{Motivation and Rationale}

Recommender systems in the news domain face the challenge of capturing both region-specific (local) and general (global) user preferences. As discussed in previous sections, users of local news sites are not exclusively interested in local content; they also value national and international news. However, traditional recommendation algorithms tend to either aggregate all user behavior globally or focus solely on local news, often missing the nuanced balance between the two. Our prior work demonstrated that localized models can outperform global models for certain news categories, for example, in \emph{Life \& Culture}, local-specific models delivered more accurate recommendations~\cite{pourashraf2022}. Nevertheless, neither purely global nor purely local models are sufficient, as both types of news consumption coexist for most users. This highlights the need for an approach that can flexibly combine local and global user preferences within a single recommender system.

To address this, we propose a hybrid recommendation architecture that integrates local and global preference models at a fine-grained level. Our approach leverages specialized models for each news category and locality (local vs. non-local), and adaptively fuses their outputs using trainable ensemble strategies. By learning how to balance the contributions of these specialized submodels, our method aims to improve the relevance and personalization of recommendations in local news contexts.



\subsection{Fusion Model Architectures}

The core of our proposed hybrid framework is the SASRec model (Self-Attentive Sequential Recommendation)~\cite{kang2018self}, which serves as the backbone for all submodels. SASRec leverages a Transformer-based self-attention mechanism to model the sequential patterns in user interaction histories, effectively capturing which past article clicks are most relevant for predicting a user's next action. Unlike simple or memory-based collaborative filtering models, SASRec can flexibly attend to both recent and distant past behaviors, making it highly effective for session-based and sequential recommendation tasks. Previous studies have shown that properly tuned SASRec models can outperform traditional session-based k-NN and other Transformer architectures such as BERT4Rec in terms of prediction accuracy~\cite{zhang2021sbert4rec}.

\paragraph{Category- and Locality-Specific SASRec Submodels.}
To capture the diversity of user interests, we partition the training data along two axes: content category (such as \emph{News}, \emph{Sports}, \emph{Life \& Culture}) and locality (local or non-local). For each combination of category and locality, we train an independent SASRec submodel. For example, within the Syracuse dataset, this results in nine specialized SASRec networks: three categories $\times$ three locality settings (local-only, non-local-only, all-content). Each submodel is trained only on interactions that match its specific segment, enabling it to specialize in that narrow domain and act as an ``expert'' for that type of content.

This design follows a mixture of experts paradigm, where each submodel provides relevance scores for candidate articles based on its area of specialization. We apply the same partitioning scheme to both the synthetic (10$\times$ Syracuse) and EB.dk datasets, using their respective category taxonomies.

\paragraph{Fusion Strategies for Combining Submodels.}
Having trained multiple specialized SASRec submodels, the next step is to combine their output into a single recommendation list. We investigate two fusion strategies:

\begin{itemize}
    \item \textbf{Neural Fusion of Independently Trained Submodels.} For each candidate article, we collect its predicted scores from all submodels, forming a feature vector of these scores. This feature vector is then passed to a small neural network---a two-layer multilayer perceptron (MLP)---which is trained to assign a final relevance score to each candidate. The MLP fusion is trained as a binary classifier, learning to produce higher scores for the actual next-clicked item. This data-driven fusion approach allows the system to adaptively weight the contributions of each submodel based on the user context and item properties, without relying on hard-coded rules or attention mechanisms.
    
    \item \textbf{Simple Ensemble Fusion (for comparison).} As a baseline, we also combine the submodel outputs using the aggregation of the mean rank. For each candidate item, we average its ranks across all submodels and recommend the items with the lowest mean rank. This ensemble method has no trainable parameters and provides a benchmark to compare against the neural fusion approach.
\end{itemize}

\paragraph{Fusion Training and Inference.}
The complete training and inference process for the neural fusion approach is summarized in Algorithm~\ref{alg:fusion}. During training, for each session and candidate set (true next-click plus negative samples), we extract submodel scores to form feature vectors and labels. The MLP fusion is then trained to optimize binary cross-entropy loss in these examples. At inference time, for a given test session, we compute submodel scores for all candidates, pass them through the fusion MLP, and recommend the top$K$ items with the highest final scores.

\begin{algorithm}[ht]
\caption{Neural Fusion of Category/Locality SASRec Submodels}
\begin{algorithmic}[1]
\REQUIRE Training data segmented by category and locality; $N$ SASRec submodels; training sessions
\STATE Train each SASRec submodel $M_i$ independently on its corresponding data segment
\FOR{each training session}
    \STATE Generate candidate items (true next item + negatives)
    \FOR{each candidate item $c$}
        \STATE Form feature vector $f_c = [M_1(\text{session}, c), M_2(\text{session}, c), ..., M_N(\text{session}, c)]$
        \STATE Label $y_c = 1$ if $c$ is the next-clicked item, else $0$
    \ENDFOR
    \STATE Store $(f_c, y_c)$ for all $c$ in current session
\ENDFOR
\STATE Train a feedforward neural network (Fusion MLP) to predict $y_c$ from $f_c$ (binary cross-entropy loss)
\STATE // Inference
\FOR{each test session}
    \FOR{each candidate item $c$}
        \STATE Get submodel scores $f_c$ as above
        \STATE Compute fusion score $s_c = \text{FusionMLP}(f_c)$
    \ENDFOR
    \STATE Recommend top-K candidates with highest $s_c$
\ENDFOR
\end{algorithmic}
\label{alg:fusion}
\end{algorithm}

\paragraph{Baselines for Comparison.}
To demonstrate the benefits of our fusion approach, we compare against three unified SASRec baselines:
\begin{itemize}
    \item \textbf{Unified All-Data SASRec:} A single model trained on the entire dataset (all categories and localities).
    \item \textbf{Unified Local-Only SASRec:} A single model trained only on local news interactions.
    \item \textbf{Unified Non-Local-Only SASRec:} A single model trained only on non-local news interactions.
\end{itemize}
These baselines allow us to quantify the advantage of fine-grained modeling and fusion over more conventional global recommendation strategies. In the next section, we discuss our evaluation methodology and experimental results.

\section{Experimental Results and Discussion}

\subsection{Datasets and Category-Locality Annotations}

We evaluated our recommendation models on two datasets, each with explicit category labels and locality annotations (or inferred locality) for every news article.

\paragraph{Synthetic Dataset (10$\times$ Syracuse).}
Our primary evaluation data set is a large-scale synthetic collection of local news. Each synthetic article is assigned one of three content categories (\emph{News}, \emph{Sports}, or \emph{Life \& Culture}) and labeled \emph{local} or \emph{non-local}, following the same schema as the Syracuse local newspaper. The proportions of categories and locality labels, as well as user-item interaction patterns, are based on detailed statistics extracted from the real Syracuse dataset. Since the original Syracuse data do not contain enough interactions for robust deep learning experiments, we use its empirical distributions to generate a synthetic dataset approximately ten times larger. This approach preserves the overall statistical structure of categories and locality, allowing us to 'stress-test' our recommendation models in a much larger, but realistically structured corpus without using real user interaction data for evaluation.

\paragraph{EB.dk (Ekstra Bladet) Danish News Dataset.}
We also include a real-world recommendation dataset from the Danish publisher \emph{Ekstra Bladet}, sometimes referred to as EB-NeRD in prior work. It supplies its own editorial categories (specific to Danish news) and extensive user-interaction logs. Since EB.dk metadata does not include explicit local vs. non-local designations, we infer locality using a large language model (LLM): each article is assigned a local or non-local label by automated content analysis. Although these labels are approximate, they allow us to apply our category--locality modeling approach in a different linguistic and cultural context.

To assign local versus non-local labels to articles in the Danish EB-NeRD dataset, we used the Llama 3 language model with the following prompt:

\begin{quote}
You are given a news article from Ekstra Bladet (in Danish) with the following details: \\
Title: \{title\} \\
Subtitle: \{subtitle\} \\
Body: \{body\} \\

Task: \\
1. Read the title, subtitle, and body of the article carefully. \\
2. Determine whether the article is about Denmark (local) or about other countries / global topics (non-local). \\
3. Provide your classification as either 'local' or 'nonlocal'. \\

\textbf{Important}: Output only the single word 'local' or 'nonlocal' with no additional text or explanation. \\

Now, please provide the answer.
\end{quote}

For each article, we replaced the placeholders with the actual content and prompted Llama 3 to return either \texttt{local} or \texttt{nonlocal}. The model output was used directly as the locality label of the article. This automated approach enabled efficient and consistent labeling across the large corpus.

\paragraph{EB-NeRD Dataset Summary.}
All statistics in this paper are derived from a subset of the public \texttt{EB-NeRD} dataset.
For our main experiments, we selected the ten largest categories. This filtered subset contains 2,942,726 interactions, 9,108 unique articles, and 10,236 users. Within this subset, 53\% of interactions involve local news. These statistics describe the portion of the dataset that we used for modeling and evaluation in this work.

For both datasets, we split the interactions into training, validation, and test sets chronologically, allowing the most recent portion to be evaluated. We ensure that category and locality distributions remain similar across the splits so that each model sees representative examples of local and non-local content during training.

\FloatBarrier
\begin{figure}[htbp]
    \centering
    \includegraphics[width=0.95\textwidth]{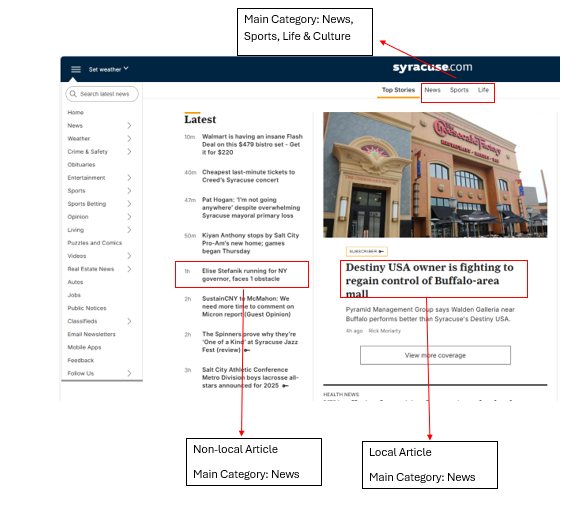}
    \caption{Homepage of Syracuse.com illustrating the organization of main categories (top navigation), subcategories (left sidebar), and an example news article. The figure highlights how articles and categories are structured on the website.}
    \label{fig:syracuse-homepage2}
\end{figure}

\subsection{Evaluation Protocol and Baseline Models}

We evaluated all model variants on held-out test sets from each dataset, using Hit Rate at various top-$K$ cutoffs ($\mathrm{HR}@K$) as our primary sequential recommendation metric. In each case, the test set comprises the most recent user interaction sessions, so the models must predict which article a user will click next based on their historical sequence.

For the fusion strategies, we ensure that the top-level attention weights or the multibranch parameters are learned during training (or a dedicated fusion training phase) using only the training and validation splits. We then compare the performance of the fused models with the baselines of a single model under identical conditions.

\paragraph{Baseline SASRec Models for Comparison}
To measure the advantage of fine-grained category--locality modeling, we compare against three simpler \emph{single-model} SASRec baselines:

\begin{itemize}
    \item \textbf{Unified All-Data SASRec}: A single SASRec model trained on the entire dataset (all categories combined, both local and non-local articles). This reflects a one-size-fits-all approach with no special treatment of categories or locality.
    \item \textbf{Unified Local-Only SASRec}: A single SASRec model trained on all local articles (across all categories), ignoring non-local content. This tests how well a model performs if it sees only local news interactions.
    \item \textbf{Unified Non-Local-Only SASRec}: A single SASRec model trained solely on non-local (global) articles across every category, omitting local content entirely.
\end{itemize}

These unified baselines allow for a direct comparison with our more elaborate, multi-model approach. They show whether the additional complexity of category-locality partitioning and fusion genuinely yields better personalization.

\subsection{Experimental Scenarios and Comparison Setups}

\paragraph{Synthetic 10$\times$ Domain.}
In the large synthetic dataset, we assess whether the relative benefits of category–locality partitioning, as previously observed on smaller datasets, persist when the number of articles and interactions is increased by an order of magnitude. This evaluation tests the ability of each approach to take advantage of more abundant data and determines whether fine-grained modeling continues to outperform a single large, unified model in a substantially scaled-up, yet realistically structured, news domain.

\paragraph{EB.dk Danish News Domain.}
We also apply our framework to the Ekstra Bladet dataset, which comes from a different language and editorial environment. Because the EB.dk category taxonomy differs from Syracuse’s three-part scheme and locality labels are inferred from LLM rather than annotated by humans, we can also gauge how robust the approach is in a scenario with potentially imperfect locality tags. Success here would indicate that the method generalizes beyond a single dataset or labeling convention.

\subsection{Evaluation Metrics and Protocol}

We report $\mathrm{HR}@K$ at multiple $K$ (e.g., 10, 20, 50) for all model variants, using only SASRec as the base model and a neural network (NN) for submodel fusion. For each variant of the model, we perform an evaluation of the synthetic test split, reporting $\mathrm{HR}@K$ at multiple $K$. This protocol allows us to directly compare the performance of global, local, non-local, category-specific, and fusion-based SASRec models.

\paragraph{Experimental Scenarios on Synthetic (10$\times$ Syracuse) Dataset.}
We compare two main modeling strategies for next-item news recommendation:
\begin{itemize}
    \item \textbf{Approach 1: Standalone Global Models.} We train a single model using \emph{all} available data, ignoring content categories and locality. We experiment with both a global SKNN and a global SASRec model.
    \item \textbf{Approach 2: Fusion of category- and locality-specific submodels} Here, we train multiple SASRec submodels, one for each segment of the data. The segments include:
    \begin{itemize}
     \item \emph{All articles within each category} \\
(\texttt{Sports\_all}, \texttt{News\_all}, \texttt{Life and Culture\_all})

        \item \emph{Local articles within each category} \\(\texttt{Sports\_local}, \texttt{News\_local}, \texttt{Life and Culture\_local})
        \item \emph{Non-local articles within each category} \\(\texttt{Sports\_non-local}, \texttt{News\_non-local}, \texttt{Life and Culture\_non-local})
        \item \emph{All local articles across categories} (\texttt{all\_local})
        \item \emph{All non-local articles across categories} (\texttt{all\_non-local})
    \end{itemize}
    These submodels are combined using either a simple ensemble (mean ranks) or a neural network (NN) fusion layer. For completeness, the same fusion procedures are also applied to the SKNN submodels.
\end{itemize}

Our goal is to determine whether leveraging fine-grained, category- and locality-aware submodels with fusion yields superior recommendation accuracy compared to training a single global model.

\begin{figure}[htbp]
    \centering
    \includegraphics[width=1\textwidth]{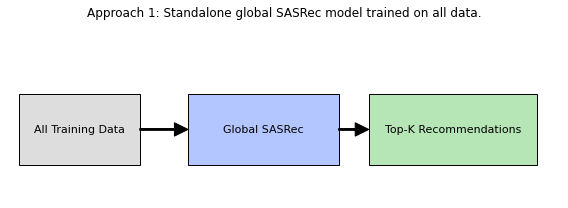}
    \caption{The baseline approach: a single global SASRec model trained on all data.}
    \label{fig:approach1}
\end{figure}

\begin{figure}[htbp]
    \centering
    \includegraphics[width=1\textwidth, height=0.4\textheight]{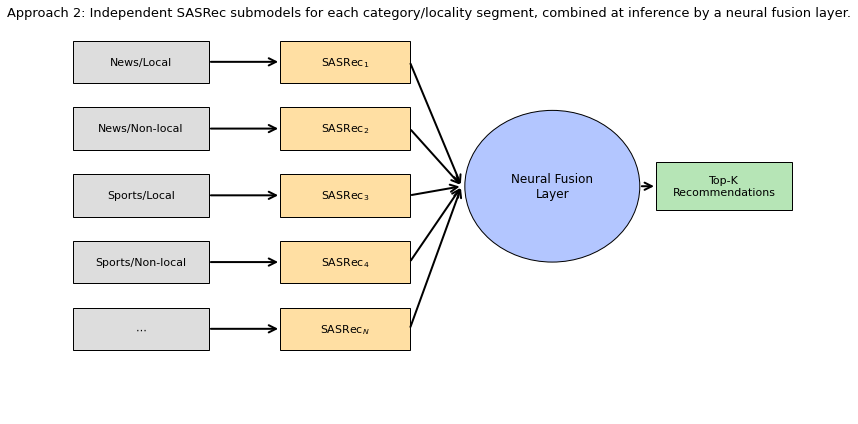}
    \caption{The proposed fusion strategy with multiple submodels and a fusion layer.}
    \label{fig:approach2}
\end{figure}

To illustrate the differences between our two main modeling strategies, Figure~\ref{fig:approach1} shows the baseline approach. In contrast, Figure~\ref{fig:approach2} shows our proposed fusion strategy.

\subsection{Results Tables}

We summarize the empirical performance of all models evaluated in terms of Hit Rate at various top$K$ thresholds ($\mathrm{HR}@K$). These results allow for direct comparison of the global baselines and our proposed fusion strategies across both the synthetic (10$\times$ Syracuse) and EB.dk datasets. For each experimental scenario, we report $\mathrm{HR}@10$, $\mathrm{HR}@20$, and $\mathrm{HR}@50$ to reflect different recommendation cut-off points.

Table~\ref{tab:synthetic-scenario-results} presents the results on the synthetic local news dataset, comparing global standalone models with ensembles of sub-models specific to the category and the locality. Table~\ref{tab:ebdk-top10} reports analogous results for the Danish news dataset EB.dk.

\begin{table}[htbp]
    \centering
    \caption{Hit Rate at K ($\mathrm{HR}@K$) for standalone global models (Approach~1) versus fused submodel ensembles (Approach~2) on the synthetic (10$\times$ Syracuse) dataset.}
    \begin{tabular}{lccc}
        \hline
        \textbf{Model} & $\mathrm{HR}@10$ & $\mathrm{HR}@20$ & $\mathrm{HR}@50$ \\
        \hline
        \multicolumn{4}{l}{\emph{Approach 1: Standalone Global Models}} \\
        \hline
        SKNN (Global)              & 0.168 & 0.288 & 0.432 \\
        SASRec (Global)            & 0.218 & 0.349 & 0.660 \\
        \hline
        \multicolumn{4}{l}{\emph{Approach 2: Fused Category/Locality Submodels}} \\
        \hline
        SKNN + Ensemble Fusion     & 0.124 & 0.236 & 0.543 \\
        SKNN + NN Fusion           & 0.142 & 0.275 & 0.614 \\
        SASRec + Ensemble Fusion   & 0.158 & 0.281 & 0.622 \\
        SASRec + NN Fusion         & \textbf{0.322} & \textbf{0.454} & \textbf{0.737} \\
        \hline
    \end{tabular}
    \label{tab:synthetic-scenario-results}
\end{table}

Table~\ref{tab:synthetic-scenario-results} shows the $\mathrm{HR}@K$ results comparing global recommendation models (Approach1, using a single SASRec base model) to ensembles of category- and locality-specific SASRec submodels fused with a neural network (Approach2).
 Among global models, the SASRec deep learning approach outperforms SKNN in all hit rate cutoffs. However, the best performance is achieved by fusing multiple SASRec submodels with a neural network (NN) layer. The SASRec + NN Fusion ensemble achieves the highest hit rates ($\mathrm{HR}@10 = 0.322$, $\mathrm{HR}@20 = 0.454$), outperforming both the global SASRec model and the ensemble-based fusion.

This shows that partitioning the dataset into fine-grained segments (by category and locality), then combining the outputs of specialized SASRec submodels using a learnable neural fusion, results in significant gains. This effect is especially pronounced at lower $K$, indicating that the fused approach produces more precise recommendations of the highest ranked. In contrast, simple ensemble fusion or SKNN-based methods provide modest improvements and sometimes trail behind the best single SASRec model.

Overall, these results highlight that category- and locality-aware modeling with neural fusion yields robust improvements over traditional global modeling. The findings support the hypothesis that the use of submodel specialization and trainable fusion layers is a powerful strategy for personalized news recommendation in large and diverse collections.

\begin{table}[htbp]
    \centering
    \caption{Hit Rate at K ($\mathrm{HR}@K$) on EB.dk (Danish) dataset for the top 10 categories (\textit{nyheder}: news, \textit{sport}: sports, \textit{krimi}: crime, \textit{underholdning}: entertainment, \textit{nationen}: the nation, \textit{penge}: money/finance, \textit{musik}: music, \textit{forbrug}: consumer, \textit{sex og samliv}: sex and relationships, \textit{ferie}: vacation/travel). Comparison between standalone global SASRec and the fusion-based SASRec approaches.}
    \begin{tabular}{lccc}
        \hline
        \textbf{Model} & HR@10 & HR@20 & HR@50 \\
        \hline
        SASRec (Global)          & 0.420 & 0.552 & 0.783 \\
        SASRec + NN Fusion       & \textbf{0.508} & \textbf{0.655} & \textbf{0.840} \\
        \hline
    \end{tabular}
    \label{tab:ebdk-top10}
\end{table}

Table~\ref{tab:ebdk-top10} presents the core results for the EB.dk dataset, evaluated on the top 10 largest news categories: \textit{nyheder} (news), \textit{sport} (sports), \textit{krimi} (crime), \textit{underholdning} (entertainment), \textit{nationen} (the nation), \textit{penge} (money/finance), \textit{musik} (music), \textit{forbrug} (consumer), \textit{sex og samliv} (sex and relationships), and \textit{ferie} (vacation/travel). English translations are included for clarity.

We compare the performance of a standalone global SASRec model with the neural fusion of specialized category/locality SASRec submodels. The fusion-based approach achieves noticeably higher hit rates at all reported values of $K$: for example, $\mathrm{HR}@10$ improves from 0.420 (global SASRec) to 0.508 (fusion), $\mathrm{HR}@20$ from 0.552 to 0.655, and $\mathrm{HR}@50$ from 0.783 to 0.840. These gains indicate that the use of fine-grained specialization and adaptive fusion leads to substantially better top-$K$ recommendation quality.

The results confirm that neural fusion is effective even in a large, real-world, multi-category news environment and that its advantages over a strong global baseline increase as the recommendation list grows. This supports our central hypothesis: \emph{explicitly modeling and fusing category- and locality-specific user preferences delivers more accurate and engaging news recommendations}.
Limitations, Future Directions, and Cross-Dataset Summary

\subsection{Limitations and Future Directions}

Although our experiments demonstrate that neural fusion of specialized SASRec submodels yields substantial gains over global baselines—especially in large, category-diverse news domains—several limitations remain. First, the performance of fusion-based methods can be sensitive to the quality of the segmentations (categories/localities), which are often based on editorial taxonomies or automated labels that may not perfectly reflect user interest clusters. Furthermore, for categories with very few interactions (for example, \textit{ sex og samliv} and \textit{ferie}), submodels trained on small data may underperform due to data sparsity, suggesting the need for more robust approaches in low-data regimes.

Looking ahead, future work should explore more dynamic segmentation and fusion strategies. For example, data-driven grouping of users or articles (using clustering or representation learning) might better capture latent preference structures than fixed editorial categories. More sophisticated fusion mechanisms, such as attention-based gating networks, could further enhance the system's ability to adaptively balance global and local signals at both the user and item level. Finally, extending the evaluation to more diverse news environments (languages, regions, and platforms) and to online user engagement metrics will be crucial to fully establishing the practical utility of the approach.

\subsection{Cross-Dataset Summary}

Across both the synthetic Syracuse dataset and the real-world Danish dataset (EB.dk), our results consistently show that integrating category- and locality-specific models with a neural fusion layer leads to the best overall recommendation accuracy, especially as the recommendation list grows larger. The gains from fusion are particularly strong in multicategory environments where user interests are heterogeneous and no single model captures all relevant signals. However, the value of fusion is context-dependent: in balanced or very large categories, global models may still provide competitive results. Future work should seek to make fusion and segmentation more adaptive, to further enhance personalization and engagement in local and global news recommender systems.

\section{Conclusion}

This paper introduced a hybrid news recommender system that explicitly fuses local and global user preference models at a fine-grained, category-specific level. Through experiments on both a large-scale synthetic local news dataset (modeled after a real U.S. city newspaper) and a real-world Danish news platform covering the ten largest editorial categories, we demonstrated that our neural fusion approach outperforms strong global baselines in predicting the next article a user will read.

The results show that neural fusion of specialized SASRec submodels yields significant improvements in top-$K$ hit rate, particularly as the size of the recommendation list increases. These gains are robust in both synthetic and real multicategory news environments, confirming that users’ interests are best captured through a balance of community-specific and general news preferences.

Despite these advances, important challenges remain. The effectiveness of our method depends on the granularity and quality of editorial categories and locality labels, and further work is needed to develop more adaptive, data-driven segmentation, and fusion strategies. In addition, practical deployment will require careful evaluation of real-world user engagement and the ability of the system to adapt to evolving news consumption patterns.

In sum, our work provides clear evidence that the integration of local and global preference signals, through neural fusion of sequential models based on categories and locality, offers a promising path to more relevant, engaging, and sustainable news personalization. We hope these findings encourage further research on hybrid recommender architectures and their impact on strengthening both local and global news ecosystems.

\bibliographystyle{unsrt}
\bibliography{references}

\end{document}